# Evidence for a Parity Broken Monoclinic Ground State in the S = ½ Kagomé Antiferromagnet Herbertsmithite

N. J. Laurita[1,2], A. Ron[1,2], J. W. Han[3], A. Scheie[4], J. P. Sheckelton[5], R. W. Smaha[5,6], W. He[5], J.-J. Wen[5], J. S. Lee[3], Y. S. Lee[5,7], M. R. Norman[8], & D. Hsieh[1,2]

[1]*Department of Physics, California Institute of Technology, Pasadena, California 91125, USA.*

[2]*Institute for Quantum Information and Matter, California Institute of Technology, Pasadena, California 91125, USA.*

[3]*Department of Physics and Photon Science, Gwangju Institute of Science and Technology, Buk-gu, Gwangju 61005, Korea.*

[4]*Institute for Quantum Information and Matter, Department of Physics and Astronomy, Johns Hopkins University, Baltimore, Maryland 21218, USA.*

[5]*Stanford Institute for Materials and Energy Sciences, SLAC National Accelerator Laboratory, Menlo Park, California 94025, USA*

[6]*Department of Chemistry, Stanford University, Stanford, CA 94305, USA.*

[7]*Department of Applied Physics, Stanford University, Stanford, CA 94305, USA.*

[8]*Materials Science Division, Argonne National Laboratory, Argonne, Illinois 60439, USA.*

**Nearest-neighbor interacting S = ½ spins on the ideal Kagomé lattice are predicted to form a variety of novel quantum entangled states, including quantum spin-liquid (SL)[1–4] and valence bond solid (VBS) phases[5,6]. In real materials, the presence of additional perturbative spin interactions may further expand the variety of entangled states[7–9], which recent theoretical analyses show are identifiable through the spontaneous loss of particular discrete point group symmetries[3,4,10]. Here we comprehensively resolve the ground state point group symmetries of the prototypical Kagomé SL candidate $ZnCu_3(OH)_6Cl_2$ (Herbertsmithite)[11] using a combination of optical ellipsometry and wavelength-dependent multi-harmonic optical polarimetry. We uncover a subtle parity breaking monoclinic structural distortion at a temperature above the nearest-neighbor exchange energy scale. Surprisingly, the parity-breaking order parameter is dramatically enhanced upon cooling and closely tracks the build-up of nearest-neighbor spin correlations, suggesting that it is energetically favored by the SL state. The refined low temperature symmetry group greatly restricts the number of viable ground states, and, in the perturbative limit, points toward the formation of a nematic $Z_2$ striped SL ground state[4] – a SL analogue of a liquid crystal.**

Spin-liquids are quantum entangled paramagnets in which classical magnetic order is precluded by virtue of geometric frustration and quantum fluctuations[12]. While these states are conventionally thought to preserve all of the symmetries inherent to the underlying crystal lattice, recent theoretical calculations have shown that the stability of certain SL phases on the Kagomé lattice may in fact hinge on the concurrent loss of discrete point group symmetries, linking the formation of gapless $Z_2$ SLs with a loss of inversion or rotational symmetries[10], chiral SLs with a loss of mirror reflection and time reversal symmetries[3,9], and gapped $Z_2$ nematic SLs with a loss of rotational symmetries[4]. This suggests that the detection of subtle changes in point group symmetries may provide a new avenue for identifying the SL ground states of Kagomé materials.

The prototypical Kagomé SL candidate $ZnCu_3(OH)_6Cl_2$ (Herbertsmithite)[11] is an excellent testbed for this approach. Its reported $R\bar{3}m$ structure consists of structurally perfect $Cu^{2+}$ (S = ½) Kagomé planes that are ABC stacked along the *c*-axis and separated by non-magnetic $Zn^{2+}$ triangular planes[13,14]. With strong nearest-neighbor antiferromagnetic interactions $J \approx 180$ K[7,15] and sufficiently perturbative Dzyaloshinskii-Moriya interactions and easy-axis exchange anisotropy $\leq 0.1J$ [7,16,17], this material is believed to well approximate the nearest-neighbor Kagomé Heisenberg antiferromagnetic model (KHAFM). A SL ground state is supported by a lack of long-range magnetic order for $T > 50$ mK[15] and further corroborated by inelastic neutron scattering[18], which reported a spin structure factor consistent with singlet-formation and the continuum of magnetic scattering intensity expected of low-energy fractionalized excitations. However, the nature of the SL ground state remains unsettled, in part due to Cu/Zn inter-site disorder[19] in which ≈15 % of the Zn inter-sites are occupied by weakly correlated[20] paramagnetic $Cu^{2+}$ ions that obscure the low temperature Kagomé response in bulk averaged probes such as heat capacity[15], magnetic susceptibility[15,21], and inelastic neutron scattering[22–24]. Consequently, a better understanding of the fundamental features, such as the symmetry of the ground state or the presence of a finite spin gap[18,20,21], would benefit from additional experimental probes.

The reported $\bar{3}m$ point group of Herbertsmithite is generated by the three-fold rotational symmetry of the Kagomé plane (Fig. 1a), a two-fold rotational axis along the *b*-axis (Fig. 1b), and spatial inversion symmetry (Fig. 1c). These symmetries must naturally be encoded in the structure of Herbertsmithite's linear and non-linear optical response tensors. The linear and third harmonic generation (THG) responses are governed by even-rank electric-dipole susceptibility tensors $\chi_{ij}^{ED}$ and $\chi_{ijkl}^{ED}$ respectively, which are non-zero for all crystal point groups. On the other hand, the leading order $\chi_{ijk}^{ED}$ tensor that governs second harmonic generation (SHG) is only non-zero in materials which do not possess an inversion center, leaving SHG to arise from either surfaces or weaker bulk electric-quadrupole $\chi_{ijkl}^{EQ}$ or magnetic dipole $\chi_{ijk}^{MD}$ processes in centrosymmetric crystals. As we will demonstrate below, examination of these three optical harmonic tensors allows for the uncovering of a subtle global distortion of the Kagomé planes of Herbertsmithite.

To track changes in the point group symmetries of Herbertsmithite, we performed temperature dependent optical rotational anisotropy (RA) measurements on double-side polished naturally grown (101) faces of plate-like Herbertsmithite single crystals (see Supplementary Note 1) using a high-speed rotating scattering plane based technique[25] (Fig. 1d). An incident laser beam of frequency ω, chosen to be resonant with $Cu^{2+}$ $d$-$d$ excitations as revealed by linear response ellipsometry measurements (Fig. 1e), is linearly polarized to be either parallel (P) or perpendicular (S) to the scattering plane before being focused to an ≈ 30 μm spot on the sample at angle of incidence Θ. The reflected or transmitted intensity of the SHG ($I^{2\omega}$) and THG ($I^{3\omega}$), also polarized to be S or P, is then isolated via optical filters and detected by a stationary two-dimensional CCD camera. By rotating the optics along the central beam axis, the RA of the harmonic intensity is mapped to a circle on the CCD, which may then be refined to determine the underlying crystal point group and non-linear generation process.

The temperature evolution of the normalized linear response intensity ($I^{\omega}$) at the relevant incident photon energies for our non-linear harmonic generation experiments is shown in Figures 2a-d. Across a wide energy range spanning the local $Cu^{2+}$ $d$-$d$ transitions, $I^{\omega}$ displays no significant temperature dependence upon cooling. Despite this, $I^{2\omega}$ (Fig. 2e) displays a striking 50-fold increase upon cooling from room temperature, first exhibiting an approximately linear increase from the noise floor followed by an enhanced onset-like behavior at $T^* = 160$ K. This temperature scale coincides with the temperature at which electron spin resonance measurements observed a build-up of nearest-neighbor spin correlations[16], the first evidence that $I^{2\omega}$ may be emblematic of the increase of spin correlations. Despite the strong temperature dependence displayed by $I^{2\omega}$, the RA-SHG patterns for all polarization configurations exhibit a uniform growth with identical temperature dependence at all scattering plane angles (Figs. 2f-i), suggesting that no change in symmetry occurs below room temperature. In contrast to $I^{2\omega}$ but reminiscent of $I^{\omega}$, $I^{3\omega}$ displays no discernible temperature dependence upon cooling (Fig. 2e). This dichotomy in the temperature evolution of $I^{\omega}$, $I^{2\omega}$, and $I^{3\omega}$ suggests a dramatic enhancement of at least one SHG process, either intrinsic to the refined $\bar{3}m$ structure or possibly signifying lower global symmetry than previously refined.

To determine the underlying non-linear process responsible for $I^{2\omega}$, we performed a global refinement of our RA data and refer to the coefficient of determination $R^2$ to quantify the quality of the global fits for each non-linear process. This analysis (Fig. 3a) reveals that neither individual nor combinations of the $\chi^{EQ}$ or $\chi^{MD}$ SHG processes permitted in the refined bulk $\bar{3}m$ structure are capable of reproducing the RA-SHG patterns, as exemplified by their exceptionally low $R^2$ values (see Supplementary Note 3). While a surface generated $\chi^{ED}$ process can reasonably reproduce the RA-SHG patterns, there are several factors which indicate a bulk origin for $I^{2\omega}$. Firstly, the magnitudes of the non-linear tensor elements of Herbertsmithite in physical units[26,27] are exceptionally large $|\chi^{ED}| \approx 1 - 30$ pm V$^{-1}$ (see Supplementary Note 4), on par with archetypal bulk SHG materials such as $\beta - BaB_2O_4$ and Quartz (Fig. 3b). Secondly, RA-SHG experiments we performed on the similar $Cu^{2+}$ Kagomé materials hexagonal Barlowite 2 and Zn-Barlowite[28] did not yield any detectable $I^{2\omega}$ (see Supplementary Note 5), despite possessing a similar *d-d* structure as Herbertsmithite and thus presumably comparable $\chi^{EQ}$, $\chi^{MD}$, and surface $\chi^{ED}$ responses. Thirdly, as we will demonstrate below, the temperature dependence of $I^{2\omega}$ scales with well-characterized global properties of Herbertsmithite. These results suggest that $I^{2\omega}$ originates from a non-linear process that cannot be explained using the refined $\bar{3}m$ structure.

Therefore, we explored the possibility of lower symmetry in Herbertsmithite by performing similar refinements in subgroups of $\bar{3}m$ (see Supplementary Note 6). Surprisingly, this analysis indicates that $I^{2\omega}$ is best reproduced by bulk $\chi^{ED}$ processes in non-centrosymmetric point groups, namely $m$, $3m$, and 2. While dynamic bulk distortions could in principle mimic the $I^{2\omega}$ of static distortions, this would require an unrealistic scenario where a parity breaking phonon mode coherently oscillates over a length scale larger than our diffraction limit (~1 μm) so as to avoid destructive interference. Therefore, in totality our results indicate that $I^{2\omega}$ originates from a bulk $\chi^{ED}$ tensor, thus revealing a static parity breaking distortion that was missed in previous structural refinements of Herbertsmithite.

To further aid in discerning the global symmetries of Herbertsmithite, we also performed RA-THG measurements. Much like the RA-SHG data, the RA-THG data are also inconsistent with the reported $\bar{3}m$ symmetry (Fig. 3c) and are instead also better reproduced by rhombohedral or monoclinic subgroups of $\bar{3}m$ (see Supplementary Note 7), further indicating a bulk distortion. To determine which point group best reproduces the entirety of our RA dataset, we then formed a weighted average of the $R^2_{SHG}$ and $R^2_{THG}$ values for each point group and non-linear process (Fig. 3d). This analysis suggests monoclinic 2 or $m$ symmetry in Herbertsmithite, indicating not only a lack of inversion symmetry but also the absence of the $C_3$ symmetry of the Kagomé plane. The fact that no change in RA is observed upon cooling implies that this monoclinic symmetry originates from above room temperature. While at first this may appear reminiscent of Herbertsmithite's parent compound $Cu_2(OH)_3Cl$ (Clinoatacamite), which is driven into a monoclinic $2/m$ phase at $T_c \approx 400$ K by Jahn-Teller active inter-site $Cu^{2+}$ ions[29], the monoclinic phase of Clinoatacamite reportedly preserves inversion symmetry. This suggests that inter-site disorder alone cannot account for the parity broken monoclinicity observed here in Herbertsmithite.

Having uncovered this monoclinic distortion in Herbertsmithite, we may now look back on previous experimental results to resolve several outstanding issues. First, although recent ESR experiments[30] were unable to ascertain the temperature dependence or parity of the low symmetry phase, they reported a minute global breaking of the $C_3$ symmetry of the Kagomé plane at $T = 5$ K. We may now understand this lower symmetry as originating from the high temperature monoclinic distortion observed here. Secondly, infra-red reflectivity measurements[31] recently reported an anomalous temperature dependent broadening of one particular $E_u$ $Cu^{2+}$ phonon that was attributed to spin-phonon coupling. Our results indicate this broadening should instead be interpreted as a subtle splitting of the two-fold degeneracy of this mode, thereby implying an $E_u$ distortion which would lower the symmetry to monoclinic 2 or $m$, exactly the point groups suggested by our RA data refinement. In this scenario, spin-phonon coupling is expected to split the two-fold degeneracy of this $E_u$ phonon by an energy proportional to the short-range spin correlator[32] $\Delta\omega \propto \langle S_i \cdot S_j \rangle$. For reasons to be outlined below, it is then appropriate to compare this phonon's linewidth to $\sqrt{I^{2\omega}}$ (Fig. 3e), which reveals a striking similarity and thus further underscores the spin correlation related origin of the temperature dependence of $I^{2\omega}$.

The mechanism[33] by which the build-up of short-range spin correlations may enhance $I^{2\omega}$ is outlined in Figure 4a. As spin correlations develop, it becomes energetically favorable for the system to readjust the positions and bonding angles of the ions that mediate the exchange interactions in an effort to reduce the magnetic energy, even at the expense of some gain in elastic energy. Without the presence of inversion symmetry, a linear coupling is permitted between these parity breaking distortions $\Delta d \propto \langle S_i \cdot S_j \rangle$ and the $\chi^{ED}$ tensor elements. Indeed, a remarkable agreement is found between $\sqrt{I^{2\omega}}$ and the predicted spin-spin correlator for the spin-1/2 KHAFM[34] (Fig. 4b), seemingly confirming the build-up of Kagomé spin-correlations as the origin for the temperature dependence of $I^{2\omega}$. We stress that the only fitting parameter in this comparison is the nearest-neighbor exchange interaction $J$ = 170 K, which is in good agreement with previous literature[7,15]. The fact that $\sqrt{I^{2\omega}}$ is well captured by the isotropic KHAFM underscores the subtlety of the monoclinic distortion, only uncovered by $I^{2\omega}$ due to its exceptional sensitivity to parity breaking distortions as $I^{2\omega} \propto \Delta d^2$, while odd harmonics scale with a much weaker $I^{\omega}, I^{3\omega} \propto \Delta d^4$ dependence for small distortions, possibly explaining their lack of exhibited temperature dependence. Most importantly, this analysis reveals that the build-up of short-range spin correlations dramatically enhances the monoclinic distortion of Herbertsmithite, i.e. the magnetic ground state further drives monoclinicity.

Having demonstrated evidence for the monoclinic distortion, we may now discuss the ramifications of this anisotropy on the potential SL ground state. Assuming the anisotropy to be perturbative, we explored the possible SL ground states which may form on a zone-centered $E_u$ distorted Kagomé lattice, as suggested by the infra-red reflectivity measurements[31]. Of all the possible $E_u$ distortions we explored (see Supplementary Notes 8 and 9), only one distortion simultaneously explains our RA data while also being a known instability of the KHAFM, a nematic $Z_2$ striped SL phase[4]. In this phase the Kagomé lattice buckles such that the $Cu^{2+}$ ions on the bowties displace along the *c*-axis, resulting in stripes of shorter Cu-Cu bonds running along one of the three formerly equivalent directions of the Kagomé lattice (Figs. 4c). Inversion symmetry is broken by contrasting distortions of adjacent Kagomé planes (Figs. 4d), resulting in a unit cell that is double the previously refined unit cell along the formerly hexagonal *c*-axis. Interestingly, while this results in an even number of Cu ions per unit cell, the number of Cu ions per Kagomé plane remains odd and therefore precludes the formation of a VBS in the limit of weak inter-plane coupling. While in principle the stripe phase is expected to exhibit two unique domains corresponding to opposite parities, both domains result in identical RA-SHG patterns because they only differ in optical phase and therefore cannot be resolved in our measurements. However, as parity domains that are smaller than our diffraction limit would exhibit SHG signals that destructively interfere, these domains must be larger than ≈ 1 μm in size, likely stabilized by long-ranged electrostatic strain fields.

In conclusion, we have presented a new route to identifying the quantum ground states of Kagomé materials through their collective spin-lattice phenomena detected via their non-linear optical response. Using this method, we have uncovered a subtle high temperature crystallographic distortion in Herbertsmithite that may ultimately select the quantum ground state, possibly resulting in the proposed nematic $Z_2$ striped SL phase[4] that is the SL analogue of a liquid crystal. While the origin of the distortion and the potential role played by inter-site disorder remain to be understood, the fact that the distortion appears to grow with the build-up of short-range spin correlations suggests a magneto-elastic mechanism, possibly similar to the "spin-Teller" driven distortions observed in frustrated pyrochlore systems[35,36] where a structural transition occurs to relieve extensive spin degeneracy. The fact that this parity breaking distortion was not observed in Barlowite 2 or Zn-Barlowite suggests the distortion and the resulting striped SL phase may be unique to Herbertsmithite, and not a generic feature of a Kagomé plane. As the striped SL phase is predicted to be $Z_2$ topologically gapped, our results corroborate previous NMR[21] and inelastic neutron scattering[20] measurements which reported a small but finite spin gap in Herbertsmithite. Additional evidence for this phase may be provided by probing for in-plane anisotropy in the thermal conductivity, as spinons are expected to be largely deconfined along the stripe direction[4]. These results highlight the exceptional sensitivity of quantum ground states to perturbations, which may result in new ground states that are not achievable in the idealized limit.

## Methods

**Ellipsometry measurements.** We performed ellipsometry measurements in the energy range of 0.74 to 6.5 eV using an M-2000, J. A. Woollam Co. spectrometer at IBS Center for Correlated Electron Systems, Seoul National University. Experiments were performed in reflection geometry at a fixed 70° angle of incidence upon the natural (101) face of a Herbertsmithite single crystal in an optical cryostat pumped down to a pressure better than $< 10^{-8}$ Torr. The sample was first cooled to base temperature with measurements performed upon heating. The complex optical conductivity was directly calculated from the ellipsometric data of $\Psi$ and $\Delta$. The $\Delta$ offset from the quartz windows were calibrated with SiO/Si wafer.

**RA-SHG and RA-THG measurements**. RA-SHG measurements were conducted in reflection geometry with an optical pulse of width $\approx$ 100 fs and center wavelength 800 nm generated by a ti:sapph amplified laser system (Coherent RegA) operating at a 100 kHz repetition rate. The angle of incidence was fixed at 10° while the scattering plane was mechanically spun about the central beam axis at $\approx$ 4 Hz. The specular reflected 400 nm SHG signal was selected using optical filters and then measured by an EM-CCD camera (Andor iXon Ultra 897). Further details regarding the RA-SHG apparatus can be found in Ref. 25. Temperature dependent RA patterns were acquired with a beam of fluence of 1.17 mJ cm$^{-2}$ and focal spot FWHM of 26 μm with 10 min exposure time. Transmission RA-SHG experiments were also conducted and show identical results to reflection experiments. RA-THG measurements were conducted in transmission geometry with an optical pulse of width $\approx$ 100 fs and center wavelength 1200 nm generated by an optical parametric amplifier operating at a 100 kHz repetition rate. The experiment was conducted at normal incidence while the polarizations of the incoming and outgoing beams were selected by manually rotating polarization optics. The transmitted 400 nm THG signal was selected and detected in an identical fashion as the RA-SHG measurements. RA patterns were acquired with a beam of fluence $\approx$ 300 μJ cm$^{-2}$ with 2 min exposure time per angle. All measurements were conducted on the same double-side polished $\sim$ 0.5 mm $\times$ $\sim$ 0.5 mm $\times$ 0.2 mm single crystal in an optical cryostat pumped down to a pressure better than $< 10^{-7}$ Torr.

**Data Availability**

The datasets generated and/or analyzed during the current study are available from the corresponding author on reasonable request.

**Additional information**

Supplementary information accompanies this paper. Correspondence and requests for materials should be addressed to D.H. (dhsieh@caltech.edu).

**Figure 1**

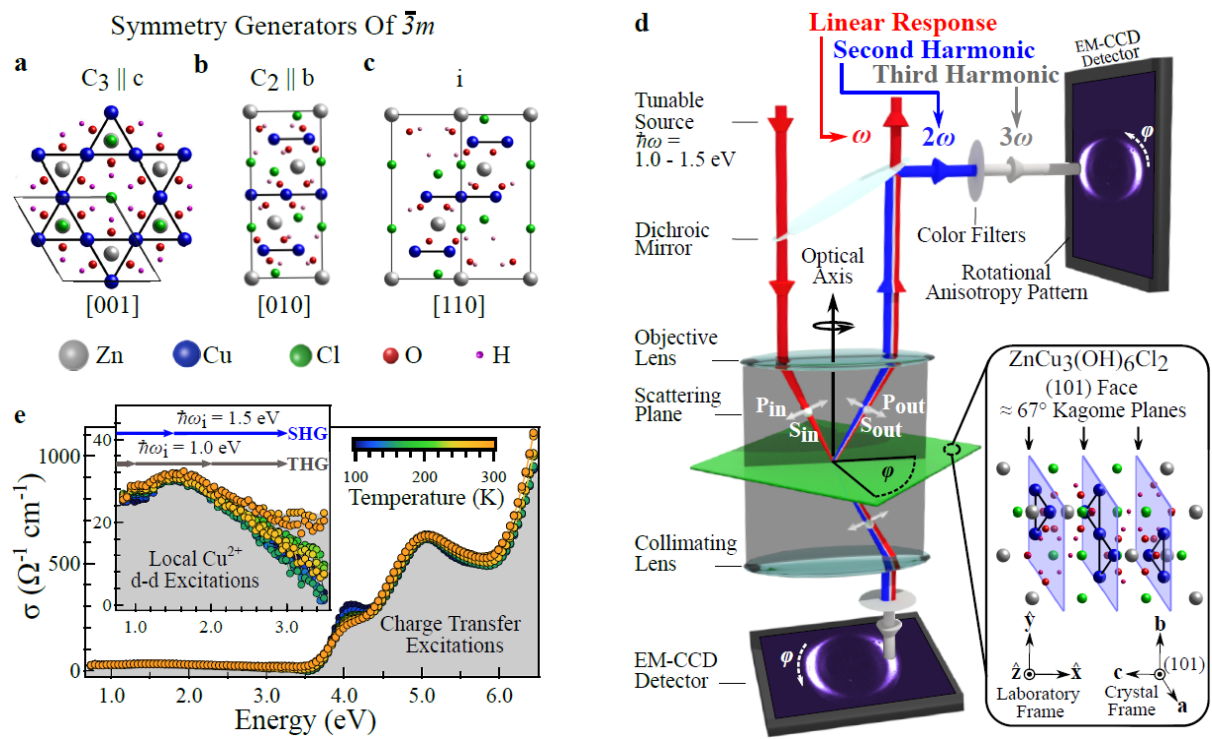

**Method for tracking point group symmetries of Herbertsmithite**. **a-c,** Structure and symmetries of Herbertsmithite in the $\bar{3}m$ point group, which is generated by **a**, a three-fold rotational axis along $\hat{c}$, **b.** a two-fold rotational axis along $\hat{b}$, and **c.** spatial inversion symmetry. **d.** Schematic of our method for measuring rotational anisotropy of the non-linear optical response. An intense laser beam of fundamental frequency ω (red) is focused on the sample at oblique incidence. The generated second (2ω, blue) and third (3ω, gray) harmonics in the reflected and transmitted beams are then isolated by optical filters and then detected by an electron-multiplying charge-coupled device camera (EM-CCD). By mechanically rotating the scattering plane about the central optical axis (black arrow), the non-linear intensity as a function of scattering plane angle φ is mapped to a circle on the detector. Beam polarizations are selected to be either parallel (P) or perpendicular (S) to the scattering plane. **e,** Optical conductivity of Herbertsmithite as measured by ellipsometry (see Supplementary Note 2). The inset displays a blown up view of the local *d-d* transitions with the energies involved in our non-linear harmonic generation experiments shown.

**Figure 2**

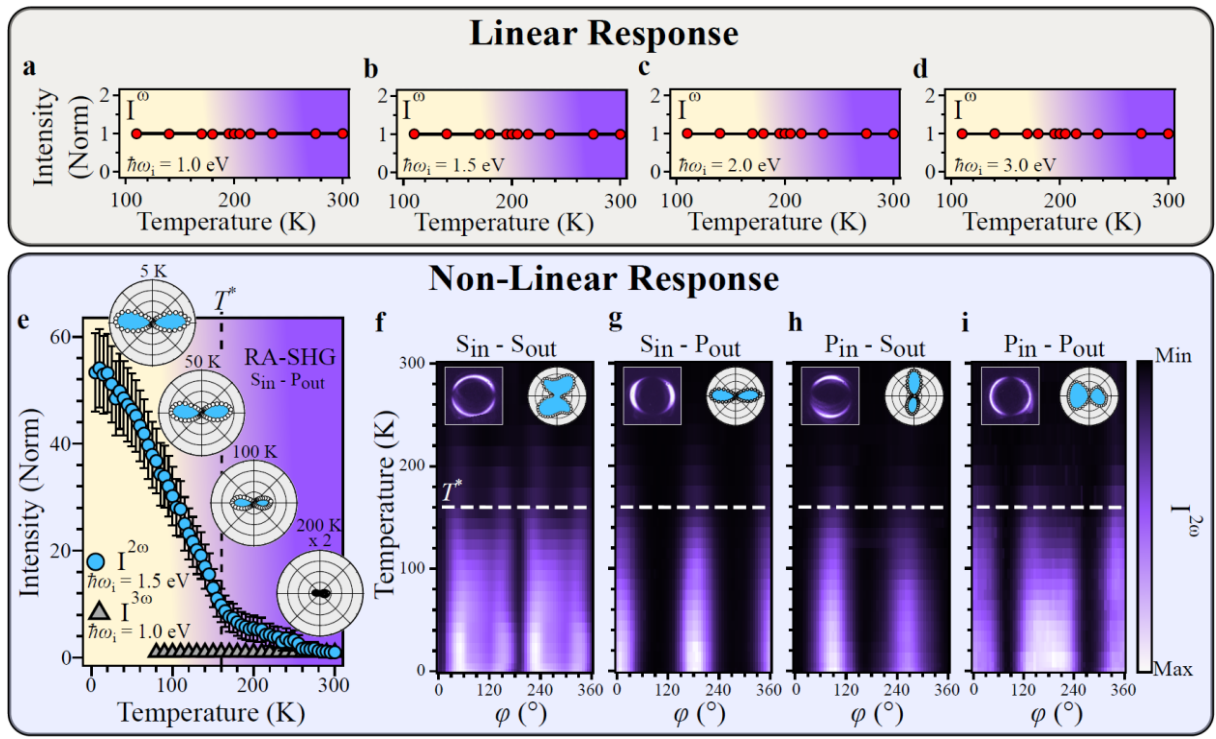

**Temperature dependent intensity of the linear and non-linear responses. a-d,** Temperature dependence of the normalized linear response intensity of Herbertsmithite ($I^{\omega}$) at incident photon energies of **a,** 1.0 eV, **b,** 1.5 eV, **c,** 2.0 eV, and **d,** 3.0 eV - the relevant energies for our non-linear optical harmonic generation experiments. No significant temperature dependence is observed. **e,** Temperature dependence of the normalized non-linear response intensity of Herbertsmithite, second harmonic ($I^{2\omega}$, blue circles) and third harmonic ($I^{3\omega}$, gray triangles). A large increase in $I^{2\omega}$ is observed at a temperature $T^* \approx 160$ K but with no corresponding change in the rotational anisotropy (insets). Image plots of the rotational anisotropy of $I^{2\omega}$ as a function of temperature in the **f,** $S_{in}$–$S_{out}$, **g,** $S_{in}$–$P_{out}$, **h,** $P_{in}$–$S_{out}$, and **i,** $P_{in}$–$P_{out}$ polarization configurations. Insets display the CCD image and corresponding polar plot of $I^{2\omega}$ at $T = 5$ K.

# Figure 3

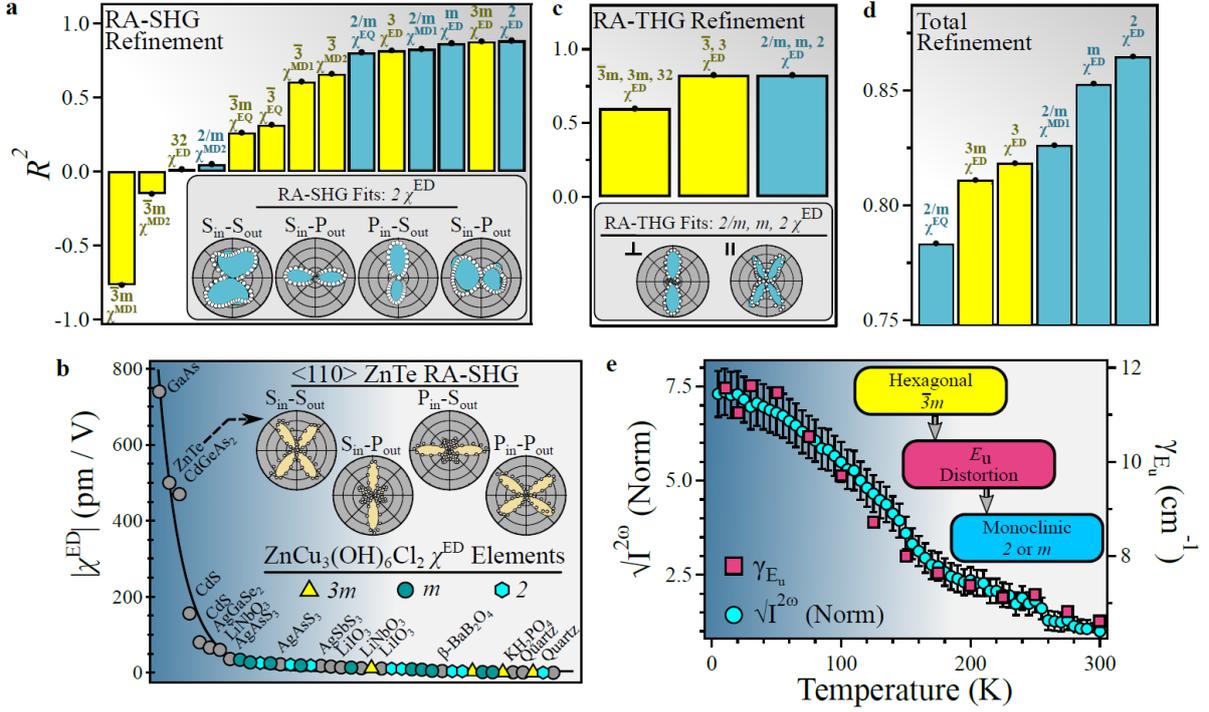

**Refinement of the non-linear optical rotational anisotropy. a,** Refinement of the $T$ = 5 K RA-SHG data as captured by $R^2$, a quantitative measure of the quality of a global fit to a given non-linear generation process. Point groups shown in yellow (teal) possess hexagonal (monoclinic) symmetry. $\chi^{ED}$ refers to the electric-dipole process $P(2\omega) = \chi^{ED}E(\omega)E(\omega)$, $\chi^{EQ}$ refers to the electric-quadrupole process $P(2\omega) = \chi^{EQ}E(\omega)\nabla E(\omega)$, $\chi^{MD1}$ refers to the magnetic-dipole process $M(2\omega) = \chi^{MD1}E(\omega)E(\omega)$, and $\chi^{MD2}$ refers to the magnetic-dipole process $P(2\omega) = \chi^{MD2}E(\omega)H(\omega)$. The RA-SHG data are best reproduced by a $\chi^{ED}$ process in the point group 2 (inset). **b,** A comparison of the extracted $T$ = 5 K $\chi^{ED}$ SHG tensor elements of Herbertsmithite for several representative point groups to those of known nonlinear materials[25], formed by referencing to the RA-SHG of ZnTe[27] (inset). Materials listed more than once possess multiple independent tensor elements. **c,** Global refinement of the $T$ = 80 K RA-THG data which were measured at normal incidence where parallel (∥) and perpendicular (⊥) refer to the relative polarizations of the incoming and outgoing beams. **d,** Total refinement of the RA-SHG and RA-THG data formed by a weighted average of their $R^2$ values as $R^2_{Tot} = (2R^2_{SHG} + R^2_{THG}) / 3$, suggesting monoclinic 2 or $m$ symmetry in Herbertsmithite. **e,** Comparison of $\sqrt{I^{2\omega}}$ to the temperature dependent splitting of the anomalous $E_u$ phonon reported by Ref. 31, as captured by its broadening linewidth (γ). Both quantities are expected to be proportional to the short-range spin correlator $\langle S_i \cdot S_j \rangle$ (see text). Inset: Any $E_u$ distortion of a $\bar{3}m$ structure would lower the symmetry to monoclinic 2 or $m$, in agreement with the total refinement of our RA data.

# Figure 4

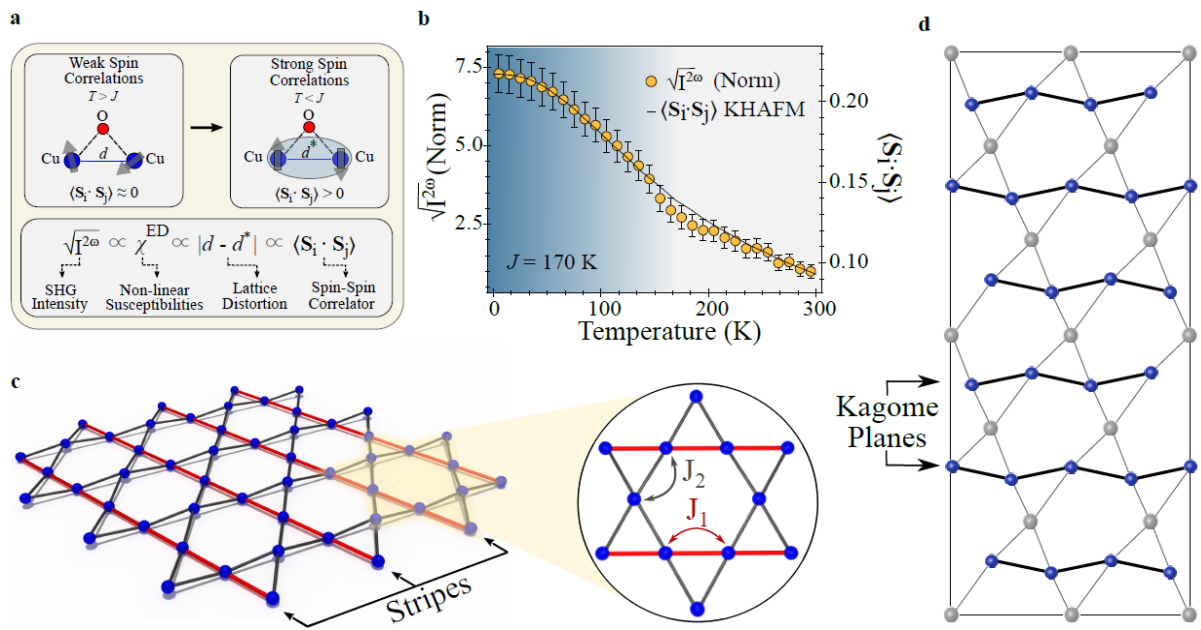

**Sensitivity of $I^{2\omega}$ to spin correlations and the proposed stripe spin-liquid ground state. a,** Schematic displaying how magneto-striction effects result in an $\sqrt{I^{2\omega}}$ that scales as the short-range spin correlator $\langle S_i \cdot S_j \rangle$. As magnetic energy increases with the development of spin correlations, the lattice may lower its overall energy by undergoing subtle distortions, denoted $\Delta d$. In the absence of inversion symmetry, these distortions, which are emblematic of the strength of $\langle S_i \cdot S_j \rangle$, in turn cause a proportional change in the $\chi^{ED}$ tensor elements and by extension $\sqrt{I^{2\omega}}$. **b,** A comparison of $\sqrt{I^{2\omega}}$ to the calculated short-range spin correlator of the nearest-neighbor Kagomé Heisenberg model with antiferromagnetic interactions (KHAFM) reported in Ref. 34. A remarkable agreement is found despite the exchange interaction $J$ = 170 K being the only fitting parameter. **c,** A zone-centered $E_u$ distorted Kagomé lattice resulting in a stripe spin-liquid ground state, a known instability of the KHAFM[4]. This phase breaks the three-fold rotational symmetry of the Kagomé plane by possessing shorter Cu-Cu bonds (red) along one of the three formerly equivalent directions of the Kagomé lattice. **d,** View of the striped phase along the hexagonal $b$ axis, such that the stripes are out of plane, displaying that the inequivalence of neighboring Kagomé planes breaks inversion symmetry. Gray and blue sphere represent $Zn^{2+}$ and $Cu^{2+}$ ions respectively.

**Acknowledgements**

This work was supported by an ARO PECASE award W911NF-17-1-0204. D.H. also acknowledges support for instrumentation from the David and Lucile Packard Foundation and from the Institute for Quantum Information and Matter, an NSF Physics Frontiers Center (PHY-1733907). N.J.L. acknowledges partial support from the IQIM Postdoctoral Fellowship. M.R.N. was supported by the Materials Science and Engineering Division, Basic Energy Sciences, Office of Science, U.S. DOE. J.W.H. and J.S.L. acknowledge support from the National Research Foundation of Korea (NRF) funded by the Ministry of Science, ICT & Future Planning under contract No. 2018R1A2B2005331. Crystal growth and characterization was performed at Stanford University and SLAC and was supported by the U.S. Department of Energy, Office of Science, Basic Energy Sciences, Materials Sciences and Engineering Division, under Contract No. DE-AC02-76SF00515. Use of the Laue machine was supported through the Institute for Quantum Matter at Johns Hopkins University, by the U.S. Department of Energy, Division of Basic Energy Sciences, Grant DE-SC-0019331. A.S was supported through the Gordon and Betty Moore foundation under the EPIQS program GBMF4532. We thank H. Changlani, S. A. Kivelson, P. A. Lee, T. Senthil, and O. Tchernyshyov for helpful conversations.

**Author contributions**

D.H., N.J.L and M.R.N conceived the experiment. N.J.L and A.R performed the non-linear harmonic generation measurements. J.W.H and J.S.L performed the ellipsometry measurements. A.S and N.J.L. performed the Laue diffraction measurements. J.P.S, R.W.S, W.H, J.J.W and Y.S.L prepared and characterized the sample. N.J.L and M.R.N analyzed the data. N.J.L, M.R.N, and D.H wrote the manuscript.

**Competing financial interests**

The authors declare no competing financial interests.

# Evidence for a Parity Broken Monoclinic Ground State in the $S=1/2$ Kagomé Antiferromagnet Herbertsmithite

N. J. Laurita,[1] A. Ron,[1] J. W. Han,[1] A. Scheie,[1] J. P. Sheckelton,[1] R. W. Smaha,[1] W. He,[1] J.-J. Wen,[1] J. S. Lee,[1] Y. S. Lee,[1] M. R. Norman,[1] and D. Hsieh*[1]

[1]*DHsieh@Caltech.edu*

## Contents

## S1. CONFIRMATION OF THE SAMPLE ORIENTATION VIA LAUE DIFFRACTION

To verify the alignment of our Herbertsmithite sample, we performed room temperature crystal diffraction measurements via a Multiwire Laboratories Laue X-ray backscatter spectrometer. Three observations confirm the naturally grown (101) orientation of the crystal surface. First, the observed diffraction pattern with the incident beam normal to the surface is consistent with scattering from the (101) plane, as evidenced by the 56° angle between the (210) and (110) Bragg peaks (Fig. S1a). Second, rotating the crystal 67° about the [010] axis yielded a diffraction pattern consistent with the expected three-fold rotational symmetry of the Kagomé plane (Fig. S1b)). Third, rotating 63° about the [0$\bar{1}$0] yielded a diffraction pattern consistent with the expected (0$\bar{1}$1) plane (not shown). These results confirm the (101) orientation of the single crystal Herbertsmithite sample investigated in this study.

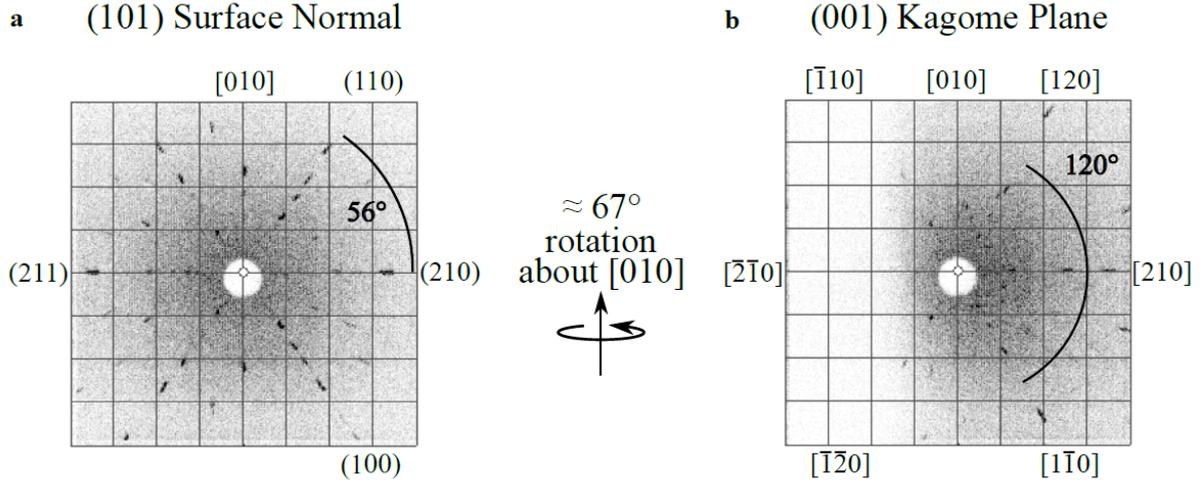

FIG. S1: Room temperature Laue diffraction measurements on the single crystal Herbertsmithite sample investigated in this study. a, Diffraction pattern obtained normal to the surface, which is consistent with the expected symmetries of the (101) plane as evidenced by the 56° angle between the (210) and (110) Bragg peaks. b, Diffraction pattern obtained by rotating the crystal 67° about the [010] direction, which displays the expected three-fold rotational symmetry of the Kagomé plane. These results confirm the (101) orientation of the sample.

## S2. PRESENTATION OF THE FULL SUITE OF ELLIPSOMETRY DATA

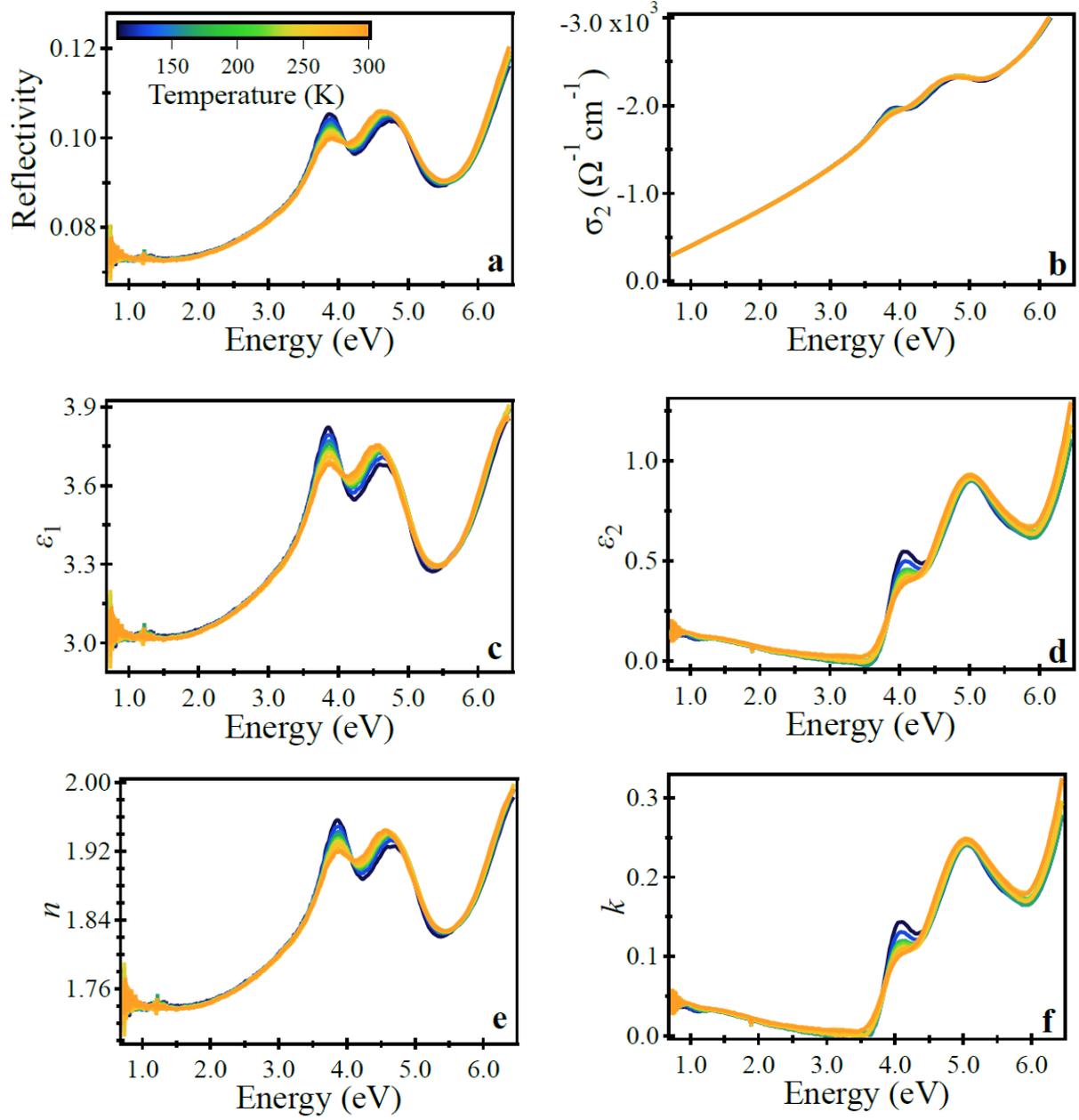

FIG. S2: Presentation of the full suite of temperature dependent ellipsometry data. a Reflectivity, b imaginary conductivity $\sigma_2$, c real permittivity $\epsilon_1$, d imaginary permittivity $\epsilon_2$, e real index of refraction $n$, f imaginary index of refraction $k$.

## S3. CATALOG OF THE RA-SHG FITS FOR SHG PROCESSES PERMITTED IN $\bar{3}$M

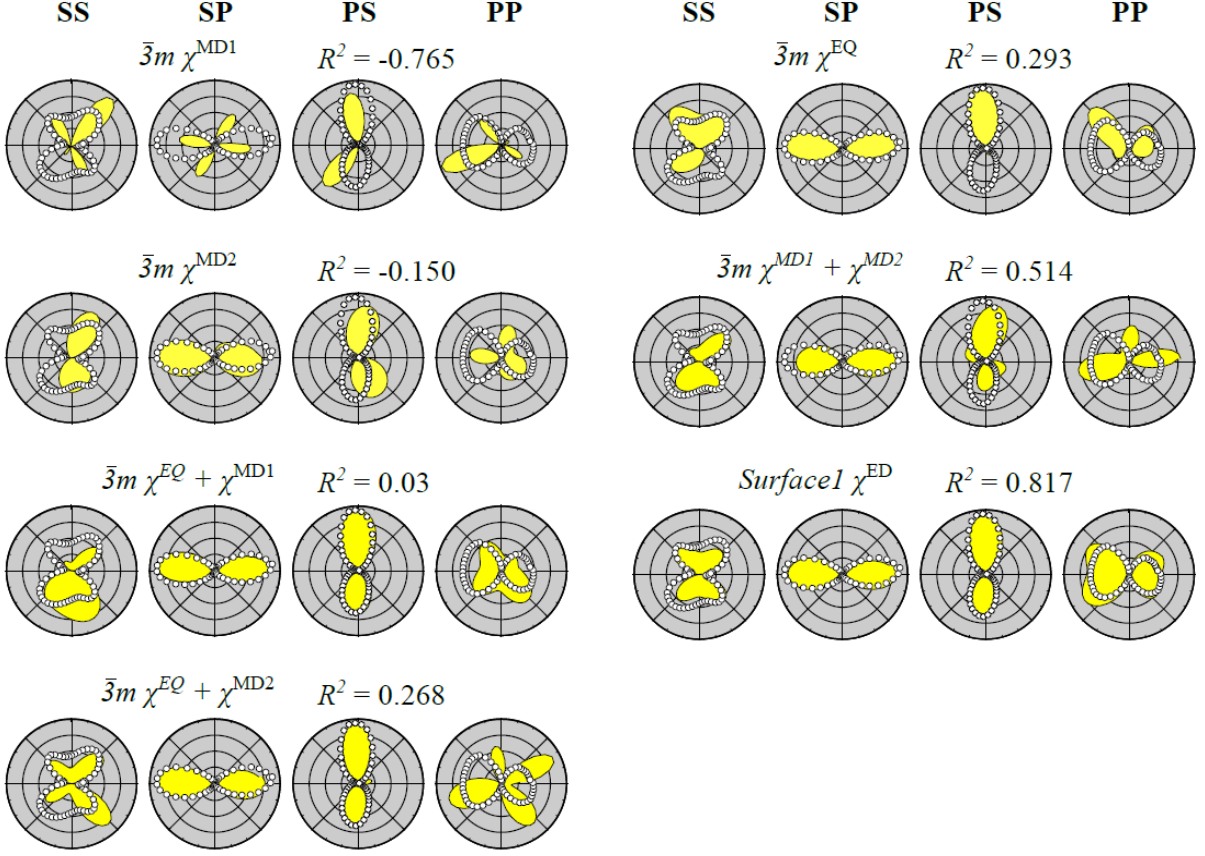

FIG. S3: Fits of the $T = 5$ K RA-SHG patterns of Herbertsmithite to individual non-linear processes and combinations thereof permitted in the previously refined $\bar{3}m$ point group. Fits are shown as black lines and yellow colored regions. Non-linear generation processes are labeled as follows: $\chi^{\text{ED}}$ represents the electric dipole process $\vec{P}(2\omega) = \chi^{\text{ED}}\vec{E}(\omega)\vec{E}(\omega)$, $\chi^{\text{EQ}}$ represents the electric quadrupole process $\vec{P}(2\omega) = \chi^{\text{EQ}}\vec{E}(\omega)\nabla\vec{E}(\omega)$, $\chi^{\text{MD1}}$ represents the magnetic dipole process $\vec{M}(2\omega) = \chi^{\text{MD1}}\vec{E}(\omega)\vec{E}(\omega)$, and $\chi^{\text{MD2}}$ represents the magnetic dipole process $\vec{P}(2\omega) = \chi^{\text{MD2}}\vec{E}(\omega)\vec{H}(\omega)$. The overall goodness of fit is captured by the $R^2$ value listed for each point group and non-linear process.

## S4. DETERMINATION OF THE $\chi^2$ TENSOR ELEMENTS OF HERBERTSMITHITE IN PHYSICAL UNITS

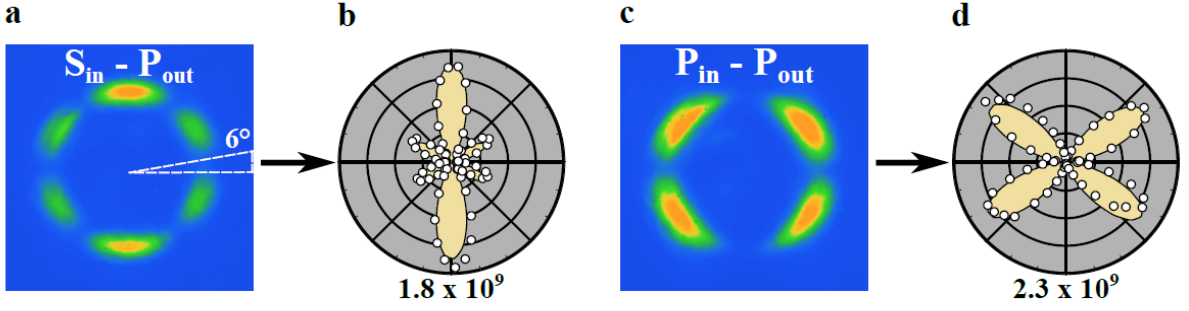

FIG. S4: **a,c**, Raw CCD images of the RA-SHG of a $\langle 110 \rangle$ oriented ZnTe single crystal at $T = 300$ K in the **a**, $S_{in}$-$P_{out}$ and **c**, $P_{in}$-$P_{out}$ polarization configurations. **b,d**, Corresponding polar plots of the SHG intensity formed by integrating the images in $6°$ segments, with the scale shown at the bottom of each polar plot.

To place the $\chi^{(2)}$ susceptibility tensor elements of Herbertsmithite in physical units, we performed RA-SHG measurements on a $\langle 110 \rangle$ oriented ZnTe single crystal under identical experimental conditions as a benchmark. The non-centrosymmetric $\bar{4}3m$ point group of ZnTe only permits one independent susceptibility tensor element $\chi_{xyz}$ in its governing electric-dipole tensor[1]. This simplicity of its non-linear optical response tensor, coupled with its well characterized nature due to its ubiquitous use in optical techniques[2], makes ZnTe a particularly well-suited standard for RA-SHG measurements.

Figures S4a & c display the EM-CCD images of the RA-SHG of ZnTe at $T = 300$ K in the $S_{in}$-$P_{out}$ and $P_{in}$-$P_{out}$ polarization configurations respectively. These images were integrated in $6°$ segments to form the rotational anisotropy patterns (Figs. S4b,d) of the total number of counts on the CCD, a quantity proportional to the SHG intensity $I^{2\omega}(\phi)$. These RA-SHG patterns were then fit to the expressions derived from $I^{2\omega}(\phi) \propto |\hat{e}_i^{2\omega}(\phi) \chi_{ijk} \hat{e}_j^\omega(\phi) \hat{e}_k^\omega(\phi)|^2 I_0^2$ where $\hat{e}$ are units vectors denoting the polarization directions and $I_0$ is the incident intensity. For light obliquely incident ($\theta = 10°$) upon the $\langle 110 \rangle$ face of ZnTe, we obtain the expressions,

$$I_{ss}^{2\omega} = 9\chi_{xyz}^2 \cos[\phi]^4 \sin[\phi]^2 \tag{S1}$$

$$I_{sp}^{2\omega} = \frac{1}{16} \cos[\frac{\pi}{18}]^2 \chi_{xyz}^2 (\cos[\phi] + 3\cos[3\phi])^2 \tag{S2}$$

$$I_{ps}^{2\omega} = (\frac{1}{2}\chi_{xyz} \cos[\frac{\pi}{18}]^2 (1 + 3\cos[2\phi])\sin[\phi] + \chi_{xyz} \sin[\frac{\pi}{18}]^2 \sin[\phi])^2 \tag{S3}$$

$$I_{pp}^{2\omega} = \chi_{xyz}^2 \cos[\frac{\pi}{18}]^2 \cos[\phi]^2 (4\sin[\frac{\pi}{18}]^4 + (\sin[\frac{\pi}{18}]^2 - 3\cos[\frac{\pi}{18}]^2 \sin[\phi]^2)^2) \tag{S4}$$

for the four independent measurement configurations. Fits to these expressions, shown in tan in Figure S4b,d and Figure 3b of the main text, are sufficient to uniquely determine $\chi_{xyz}$ in units intrinsic to the experiment. By then comparing this value to the known experimental value $\chi_{xyz} \approx 500$ pm/V, we may derive the proportionality constant to convert susceptibility tensor elements into physical units. Repeating this process for Herbertsmithite and scaling the extracted independent susceptibility tensor elements by the same proportionality constant allows for their conversion into physical units.

To complete the comparison, we must also account for the difference in Fresnel coefficients between these two materials. We refer to the relation derived by Bloembergen and Pershan[3],

$$\chi_R^{(2)} = \frac{2\chi^{(2)}}{(\epsilon^{1/2}(2\omega) + (\epsilon^{1/2}(\omega))((\epsilon^{1/2}(2\omega) + 1)(n(\omega) + 1)} \tag{S5}$$

where $\chi_R^{(2)}$ refers to the experimental value of $\chi^{(2)}$ measured in reflection geometry, $\epsilon$ is the permittivity of the sample, and $n$ is the index of refraction. Inserting the relevant values for Herbertsmithite obtained from our ellipsometry measurements into this expression and repeating the process with reported values for ZnTe[4] reveals the non-linear susceptibilities of Herbertsmithite would be measured to be $\approx$ 3x weaker in reflection geometry due to the difference in Fresnel coefficients alone. This additional scaling factor was accounted for in our analysis.

## S5. COMPARISON OF THE RA-SHG OF HERBERTSMITHITE TO THAT OF BARLOWITE 2 AND ZN-BARLOWITE

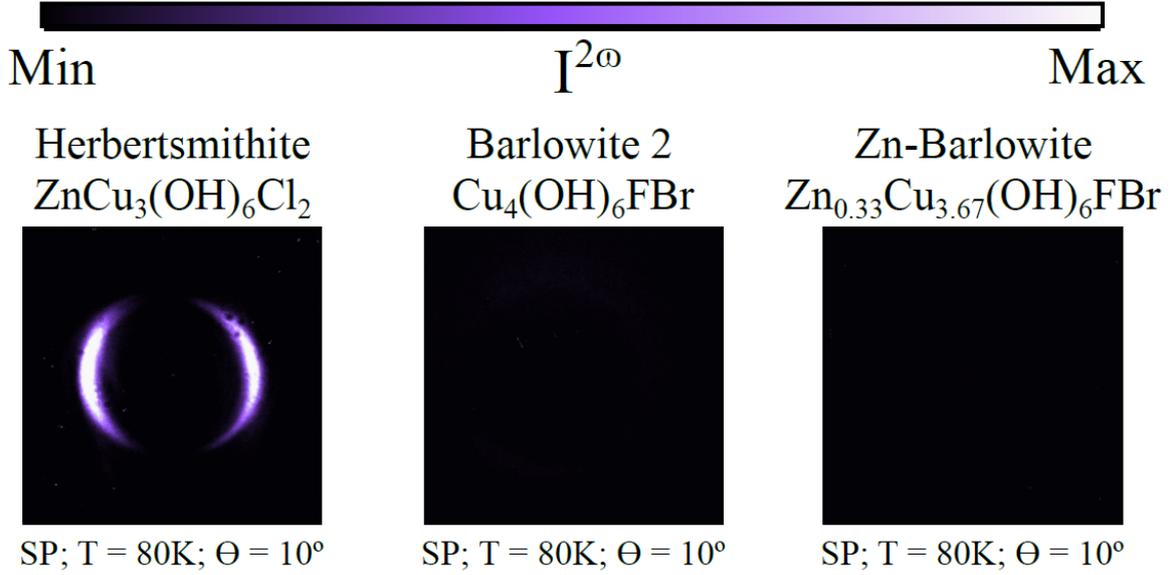

FIG. S5: Raw CCD images of the RA-SHG intensity of **a** (101) Herbertsmithite, **b** [001] Barlowite 2, and **c** [001] Zn-Barlowite[5] acquired under identical experimental conditions. Experiments were conducted in reflection geometry with 800 nm light of fluence 1.17 mJ cm$^{-2}$ obliquely incident at $\theta = 10°$ upon the sample surface at $T = 80$ K. CCD images were acquired with 10 minutes of exposure time for all samples. No discernible SHG intensity was detected in Barlowite 2 or Zn-Barlowite, despite these compounds possessing similar $d-d$ transitions in the infra-red. This further suggests that a $\chi^{ED}$ process is responsible for the observed RA-SHG of Herbertsmithite.

## S6. CATALOG OF THE RA-SHG FITS FOR SHG PROCESSES PERMITTED IN SUBGROUPS OF $\bar{3}m$

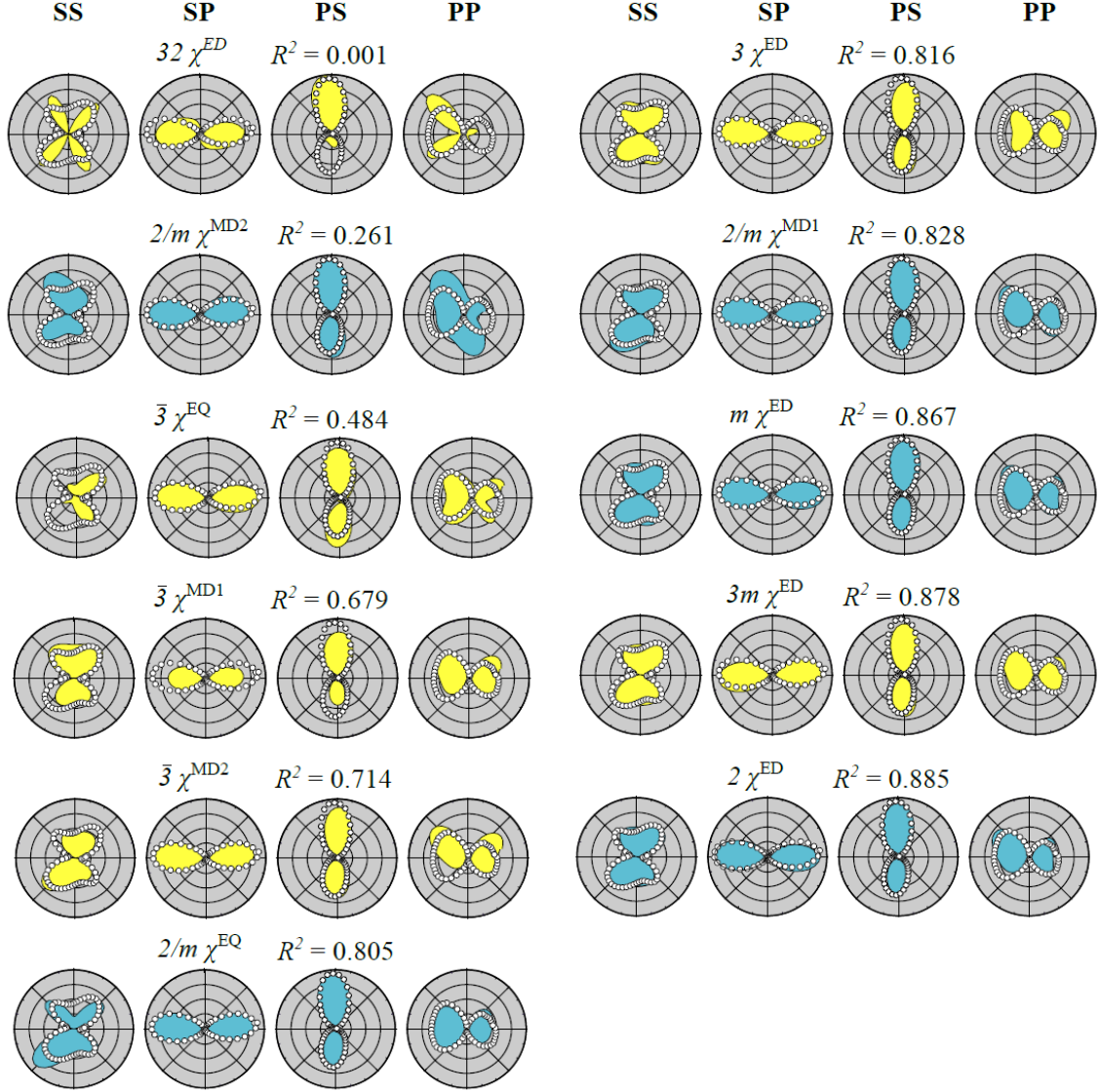

FIG. S6: Fits of the $T = 5$ K RA-SHG patterns of Herbertsmithite to non-linear generation processes permitted in subgroups of $\bar{3}m$. Fits are shown as black lines and colored regions where yellow and blue denote rhombohedral and monoclinic point groups respectively. Non-linear generation processes are labeled as follows: $\chi^{\text{ED}}$ represents the electric dipole process $\vec{P}(2\omega) = \chi^{\text{ED}}\vec{E}(\omega)\vec{E}(\omega)$, $\chi^{\text{EQ}}$ represents the electric quadrupole process $\vec{P}(2\omega) = \chi^{\text{EQ}}\vec{E}(\omega)\nabla\vec{E}(\omega)$, $\chi^{\text{MD1}}$ represents the magnetic dipole process $\vec{M}(2\omega) = \chi^{\text{MD1}}\vec{E}(\omega)\vec{E}(\omega)$, and $\chi^{\text{MD2}}$ represents the magnetic dipole process $\vec{P}(2\omega) = \chi^{\text{MD2}}\vec{E}(\omega)\vec{H}(\omega)$. The overall goodness of fit is captured by the $R^2$ value listed for each point group and non-linear process.

## S7. CATALOG OF THE RA-THG FITS

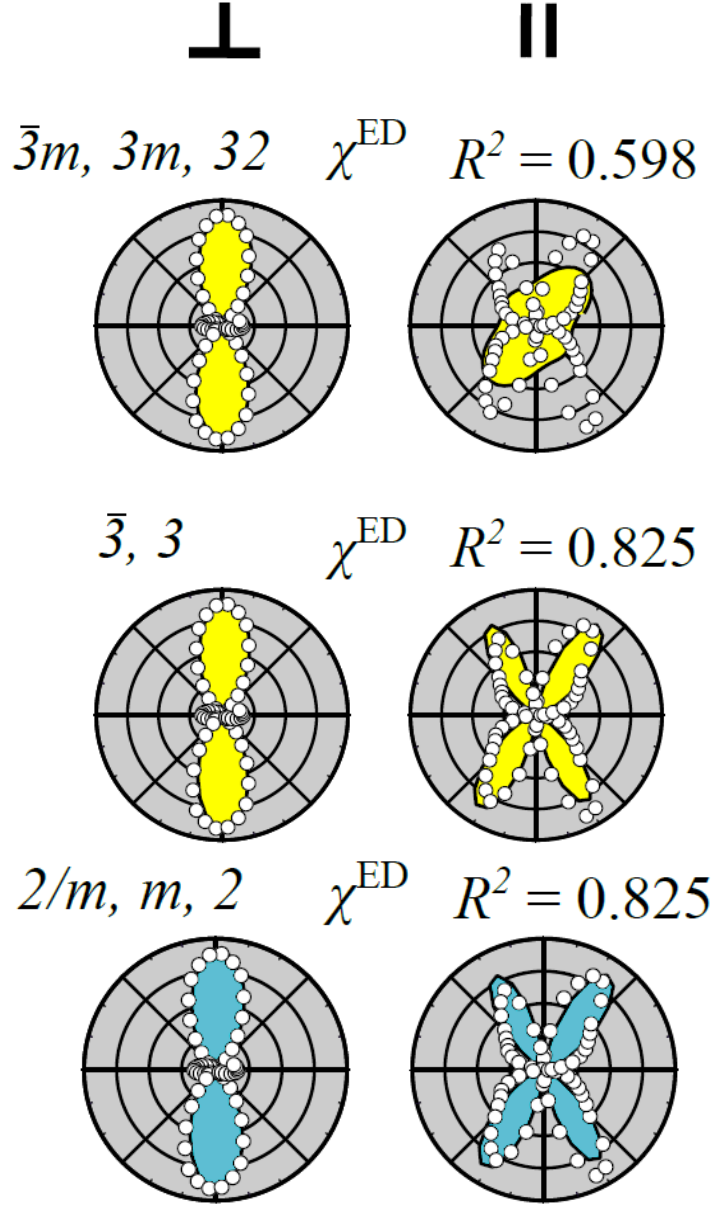

FIG. S7: Fits of the $T = 80$ K RA-THG patterns of Herbertsmithite to non-linear generation processes permitted in the $\bar{3}m$ point group and its subgroups. Data was acquired at normal incidence where the two independent measurement configurations are distinguished by the relative orientations, either parallel $\parallel$ or perpendicular $\perp$, of the incoming and outgoing light polarizations. Fits are shown as black lines and colored regions where yellow and blue denote rhombohedral and monoclinic point groups respectively. $\chi^{\text{ED}}$ represents the electric dipole process $\vec{P}(3\omega) = \chi^{\text{ED}} \vec{E}(\omega) \vec{E}(\omega) \vec{E}(\omega)$. The overall goodness of fit is captured by the $R^2$ value listed for each point group and non-linear process.

## S8. VALENCE BOND ORDERINGS OF OTHER VIABLE MONOCLINIC GROUND STATES

In the main text, we argued that Herbertsmithite is pushed deeper into a preexisting monoclinic phase by the build up of spin correlations as the temperature is reduced. This was supported in part by the refinement of the totality of our low temperature RA data, where it was found that the best fits were provided by the monoclinic point groups $m$ and 2. This is further supported by the subtle splitting of one particular $Cu^{2+}$ $E_u$ mode upon cooling as observed in the infra-red reflectivity study of Sushkov et al.[6] Indeed, a rudimentary group theory analysis reveals that any $E_u$ distortion of a $\bar{3}m$ structure would lower the point group symmetry to either $m$ or 2, exactly the point groups suggested by our RA refinement.

This led us to explore ground states which may result from zone-centered $E_u$ distortions in Herbertsmithite, from which we highlighted a potential stripe spin liquid phase. However, this is not the only viable ground state, as any $E_u$ distortion would be consistent with the symmetries we observed in Herbertsmithite. Figures S8a-c display additional monoclinic ground states which result from distinct zone-centered $E_u$ distortions. For simplicity, we only show displacements of copper ions; in reality, oxygen ion displacements will be equally important. The right panels show distortions (red arrows) of the Cu ions (blue) in the $\bar{3}m$ unit cell. The left panels show the resultant distorted Kagomé plane where Cu-Cu bonds are drawn as black, red, and green lines to denote their lengths from shortest to longest. Figure S8d displays similar diagrams for the stripe phase discussed in the main text. Apart from being the only identified $E_u$ distortion that is consistent with a known instability of the Kagomé lattice, this ground state is also particularly simple, possessing only two unique Cu-Cu bonds while the other ground states possess three.

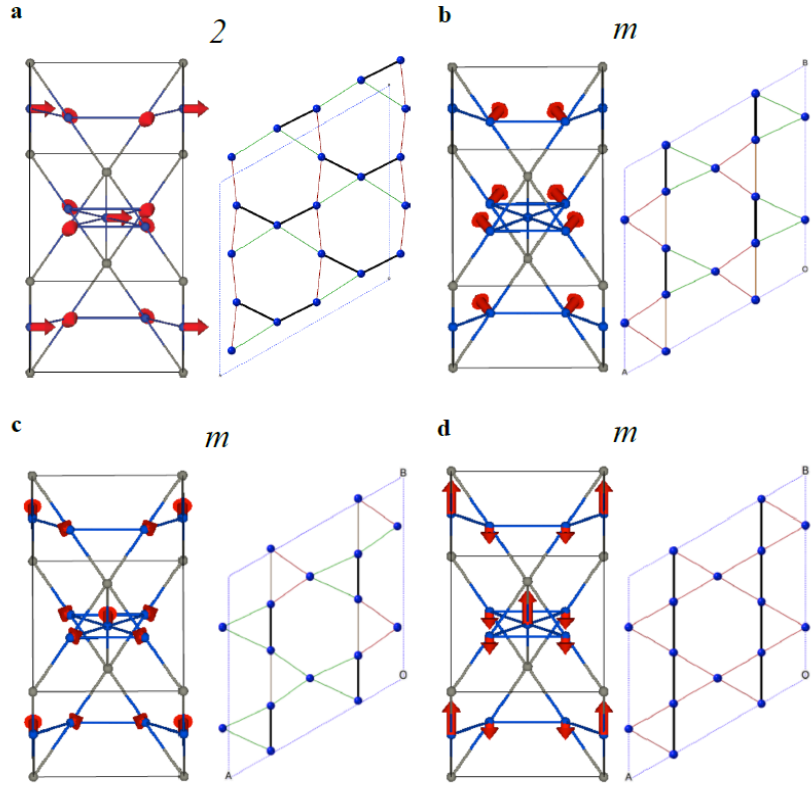

FIG. S8: Potential monoclinic ground states of Herbertsmithite which result from zone-centered $E_u$ distortions of the previously refined $\bar{3}m$ structure. Blue and gray spheres represented Cu and Zn atoms respectively. Red arrows denote Cu distortions from the $\bar{3}m$ structure. The right panel of each figure shows a single Kagomé plane where Cu-Cu bonds are drawn as black, red, and green lines to denote their lengths from shortest to longest. The phases shown in **a-c** possess three bonds of different length. However, the stripe phase (**d**) possesses only two. Distortions were generated using ISODISTORT[7]. Plots with distortion arrows were generated by VESTA[8]. Bond plots were generated by CrystalMaker.

## S9. ON THE POSSIBILITY OF NON-ZONE CENTERED $E_u$ DISTORTIONS

In the main text, we explored the potential quantum ground states that may form on a zone-centered $E_u$ distorted Kagomé lattice. This was motivated by the observed anomalous behavior at the zone-center of one particular $E_u$ phonon by infra-red reflectivity[6]. However, non-zone-centered distortions would in principle also be consistent with the monoclinic symmetry observed in our experiments. For instance, the monoclinic phase may be obtained by instead condensing a zone-boundary $F$ phonon, as occurs in Herbertsmithite's parent compound Clinoatacamite[9]. However, these finite momentum distortions do not preserve translational symmetry, resulting in an in-plane doubling of the unit cell and therefore an even number of $Cu^{2+}$ atoms per plane in the unit cell. Thus, these distortions would be expected to destabilize the spin liquid ground state and instead result in the formation of a valence bond solid.

---